\documentclass[sigplan,nonacm]{acmart}

\usepackage{fancybox}
\usepackage{pifont}

\newcommand{\systemname}{\textsc{Entwine}}

\newcommand{\motivationinsight}[3]{%
  \par\addvspace{0.6\baselineskip}%
  \noindent\begingroup
  \setlength{\fboxsep}{5pt}%
  \cornersize*{7pt}%
  \Ovalbox{%
    \begin{minipage}{\dimexpr\linewidth-2\fboxsep-1.6pt\relax}
      \itshape
      {\rightskip=0pt plus 2em\relax
       \pretolerance=-1\hyphenpenalty=50\exhyphenpenalty=50\relax
       \textbf{\ding{72}\,Insight #1: #2.}} #3
    \end{minipage}%
  }%
  \endgroup\par\nobreak\addvspace{0.6\baselineskip}%
}

\newcommand{\gemmrs}{GEMM--Reduce\-Scatter}

\newcommand{\tpeightOursSpeedup}{1.232$\times$}

\newcommand{\tpeightVsTorchMax}{1.433$\times$}

\newcommand{\tpeightFOLatencyDelta}{8.88\%}
\newcommand{\tpeightAsyncLatencyDelta}{6.17\%}
\newcommand{\tpeightFLUXLatencyDelta}{2.96\%}
\newcommand{\tpeightOursWins}{13/15}

\newcommand{\tpeightOverlapEfficiency}{0.946}

\newcommand{\tpeightOracleRatio}{1.0185$\times$}

\newcommand{\exposedShareRange}{11.0--40.2\%}
\newcommand{\exposedShareMedian}{23.3\%}

\newcommand{\scalingFOtwo}{1.0204$\times$}

\newcommand{\scalingFOeight}{1.0975$\times$}
\newcommand{\scalingTorchtwo}{1.1484$\times$}

\newcommand{\scalingTorcheight}{1.2318$\times$}

\newcommand{\envelopeFO}{1.19--2.37}
\newcommand{\envelopeFLUX}{9.48--18.96}

\newcommand{\modelPointErr}{3.0\%}
\newcommand{\modelMedianErr}{0.37\%}
\newcommand{\layerFourKFluxLead}{5.8\%}
\newcommand{\layerEightKFluxGain}{8.1\%}
\newcommand{\layerSixteenKFluxGain}{6.5\%}
\newcommand{\granularityKtwoSpeedup}{1.439$\times$}
\newcommand{\granularityKfourSpeedup}{1.199$\times$}
\newcommand{\granularityKeightSpeedup}{1.132$\times$}
\newcommand{\tileOrderLongK}{2048}
\newcommand{\tileOrderShortK}{512}
\newcommand{\tileOrderLongSpeedup}{1.267$\times$}
\newcommand{\tileOrderShortSpeedup}{1.277$\times$}
\newcommand{\tileOrderLongTailReduction}{97.6\%}
\newcommand{\tileOrderShortTailReduction}{98.5\%}
\newcommand{\sharedPartitionLongSpeedup}{1.185$\times$}
\newcommand{\sharedPartitionShortSpeedup}{1.066$\times$}

\newcommand{\Breq}{B_{\mathrm{req}}}

\newcommand{\gpuname}{A800-SXM4-80GB}

\newcommand{\cudaver}{12.1}
\newcommand{\torchver}{2.5.1+cu121}
\newcommand{\cublasver}{12.1.3.1}
\newcommand{\ncclver}{2.21.5}
\newcommand{\cutlassver}{08185b9c}
\newcommand{\flashoverlapcommit}{38fe6a3}
\newcommand{\fluxcommit}{ffb34a73}
\newcommand{\warmupcount}{120}
\newcommand{\iterscount}{200}
\newcommand{\corrAtol}{0.1}
\newcommand{\corrRtol}{0.05}

\newcommand{\waveOperatorRange}{19--114}
\newcommand{\waveProjectionSweepRange}{1.2--19}
\newcommand{\waveSuiteMeet}{19}

\title{Entwine: Coordinating Tiled Computation and Fine-Grained Communication across GPUs}

\newcommand{\affsep}{,}
\newcommand{\pdfauthors}{Kai Ma, Quanfeng Lv, Jingguo Ge, Bowei Dai, Kefan Ruan}
\author[Ma et al.]{%
  \texorpdfstring{%
    Kai Ma\textsuperscript{1\affsep2\affsep3}\quad
    Quanfeng Lv\textsuperscript{1\affsep2\affsep3}\quad
    Jingguo Ge\textsuperscript{1\affsep2\affsep3\affsep*}\quad
    Bowei Dai\textsuperscript{4\affsep*}\quad
    Kefan Ruan\textsuperscript{1\affsep2\affsep3}}%
  {\pdfauthors}}
\affiliation[obeypunctuation=true]{%
  \institution{%
    \begin{tabular}{c}
      \textsuperscript{1}State Key Laboratory of Cyberspace Security Defense\\
      \textsuperscript{2}Institute of Information Engineering, Chinese Academy of Sciences\\
      \textsuperscript{3}University of Chinese Academy of Sciences\\
      \textsuperscript{4}Institute of Microelectronics, Chinese Academy of Sciences\\[2pt]
      \href{mailto:makai@iie.ac.cn}{makai@iie.ac.cn}\quad
      \href{mailto:lvquanfeng@iie.ac.cn}{lvquanfeng@iie.ac.cn}\quad
      \href{mailto:gejingguo@iie.ac.cn}{gejingguo@iie.ac.cn}\\
      \href{mailto:daibowei@ime.ac.cn}{daibowei@ime.ac.cn}\quad
      \href{mailto:ruankefan@iie.ac.cn}{ruankefan@iie.ac.cn}\\
      \textsuperscript{*}Corresponding authors
    \end{tabular}}
  \city{}
  \country{}}

\begin{document}

\begin{abstract}
Modern high-performance GPU computations partition tensors into tiles to
exploit data reuse and parallelism. Individual tile computations complete
earlier than the full tensor computation, creating opportunities to overlap
computation and communication. However, a mismatch between computation
and communication progress can limit these opportunities. Communication
stalls when no data is ready, and may lag when data arrives in bursts.
Communication can also slow computation by consuming shared
resources, offsetting the benefits of overlap.

We present Entwine, which coordinates tile computation order,
fine-grained communication, and SM resource allocation to minimize overall
completion time. Entwine reorders tile
computation to produce data for communication at a more regular pace.
Entwine couples this schedule with fine-grained SM-based communication to
process tile results with low latency and low overhead. Since the communication kernel also consumes SM resources, Entwine
coordinates their allocation to balance communication progress against
computation slowdown. Across representative tensor-parallel
LLM workloads, Entwine achieves a geomean speedup of \tpeightOursSpeedup{} (up to
\tpeightVsTorchMax{}) over cuBLAS+\allowbreak NCCL, and outperforms
state-of-the-art overlap baselines by 3.1--9.8\% in geomean.
We will open-source our implementation upon publication.

\end{abstract}

\maketitle

\section{Introduction}
\label{sec:introduction}

Large language models are often distributed across multiple GPUs to meet
their memory and computational demands~\cite{megatron-sc21,torchtitan,alpa}.
Tensor parallelism partitions the computation of individual layers across
devices, but introduces communication to exchange and combine intermediate
results~\cite{mesh-tensorflow,gshard,gspmd}. This communication can account for a substantial fraction of
execution time and limit the benefit of parallel
computation~\cite{taccl}. Reducing communication overhead is
therefore essential to efficient multi-GPU execution~\cite{sccl}.

\begingroup
\emergencystretch=1em
Overlapping computation and communication can hide this overhead, but
data dependencies complicate their concurrent execution. When
communication depends on the output of a computation, conventional
execution waits for the entire output before starting
communication. High-performance matrix multiplication
(GEMM) kernels, however, compute output in smaller regions, or
tiles~\cite{volkov-gemm,triton},
that finish at different times. These completed tiles
provide opportunities to advance communication while the remaining
output is still being computed.
\par
\endgroup

Prior work enables dependent computation and communication to overlap
through decomposition, signaling, and kernel fusion~\cite{domino,t3,fused-collective}.
CoCoNet pipelines communication chunks with GEMM, while FlashOverlap
releases completed waves or wave groups to library
collectives~\cite{coconet,flashoverlap}.
These approaches amortize communication overhead, but tiles completed
early must wait for the rest of their communication group.
FLUX enables tile-level transfers through GEMM epilogue
fusion~\cite{flux}.
In this execution pattern, output transfers are issued as part of tile
computation, coupling communication parallelism to the GEMM execution
structure.
This coupling restricts how independently the two activities can be
scheduled and provisioned~\cite{tilelink}.
Comet instead assigns computation and communication to specialized
blocks and profiles their resource split~\cite{comet}.
However, a resource split chosen for an entire execution may not match
the changing balance between computation and communication demand within
that execution.
These execution choices constrain both when completed output can be
communicated and how communication competes with subsequent computation.

\begin{figure}[t]
  \centering
  \includegraphics[width=\columnwidth]{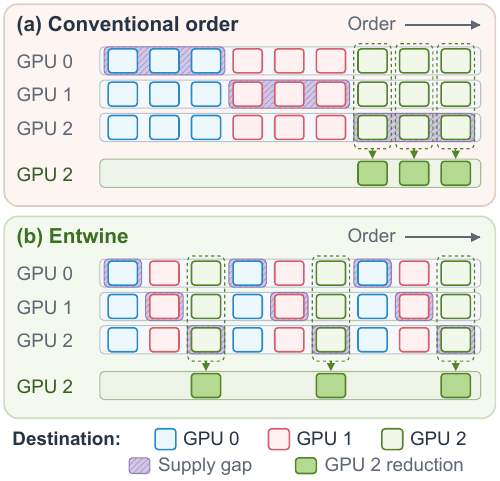}
  \caption{Tile computation order and its effect on communication in
  GEMM--ReduceScatter. Colors denote destination GPUs; purple bands mark
  locally consumed tiles. Interleaving disperses locally retained tiles,
  producing communication data at a more regular pace. Dashed lines
  connect contributions to GPU~2's reductions.}
  \Description{Two panels show three GPUs computing the same nine tile
  contributions in different logical orders. Each output partition contains
  three tiles. The top panel computes all tiles for GPU 0, followed by
  those for GPU 1 and GPU 2. The lower panel interleaves the partitions.
  Purple bands identify each producer's locally retained contributions:
  one three-tile run per GPU above and three one-tile runs below. Dashed
  columns connect the three contributions to each GPU 2 tile. A dedicated
  GPU 2 reduction swimlane in each panel shows the same three tile reductions,
  clustered late in the upper order and spread across the lower order.
  Legends are placed below both panels.
  Other destinations follow the same dependency rule. The drawing does not
  depict measured completion times or transfer and reduction durations.}
  \label{fig:intro-abstraction}
\end{figure}

These constraints lead to two coupled challenges. First, computation must
sustain the supply of tiles for communication, and communication must
process those tiles promptly. The production order must therefore be
coordinated with tile-level communication, so that early outputs
do not wait for unrelated producer blocks. Second, creating and
exploiting these opportunities must reduce end-to-end latency despite
the cost imposed on computation. Reordering can reduce data reuse, while
concurrent transfers and reductions compete for SM resources and memory
bandwidth. Both can slow subsequent tile production. A more regular
supply or a shorter communication tail is therefore valuable only when
its benefit exceeds the accompanying computation slowdown.

We present \systemname{}, which coordinates tile production, tile-level communication, and shared-SM resources to overlap computation with collective communication. We realize this design for GEMM--ReduceScatter as a concrete and widely used instance. \systemname{} interleaves tile computation across output partitions, distributing locally retained tiles among those requiring inter-GPU transfer (Figure~\ref{fig:intro-abstraction}, bottom). This order supplies communication inputs at a more regular pace. To exploit this supply promptly, \systemname{} releases and communicates output tiles individually. Each tile reduction requires the corresponding contributions from all ranks. Its input dependencies do not extend to a larger group of output tiles.

\systemname{} coordinates this tile-level execution in a shared SM pool
to balance communication progress against computation slowdown. One
computation kernel and one communication kernel execute concurrently,
without a separate launch for each tile. As blocks finish, their
resources become available to pending work from either kernel. A
communication budget bounds the number of concurrent communication
blocks, balancing the processing of completed tiles against the
production of subsequent ones. We jointly select computation and
communication configurations through offline profiling of end-to-end
latency. This selection accounts for both the overlap gained and the
computation slowdown, instead of targeting GEMM throughput or the
communication tail in isolation.

This paper makes the following contributions:
\begin{itemize}
  \item We present \systemname{}, which coordinates tile
  production, tile-level communication, and shared-SM execution to reduce
  overall completion time.

  \item We introduce an interleaved production order coupled with
  independent tile-level communication to sustain data supply and process
  completed output without cross-block assembly waits.

  \item \begingroup\emergencystretch=1em
  We coordinate computation and communication in a shared SM pool
  through a communication budget selected for end-to-end latency. We
  further relate achieved GEMM throughput to the average communication
  bandwidth needed to keep pace with production.
  \par\endgroup

  \item We evaluate Entwine across tensor-parallel workloads and GPU counts. On our main suite, it achieves a 1.232$\times$ geomean speedup over cuBLAS+NCCL, with a maximum of 1.433$\times$, and outperforms state-of-the-art overlap baselines by 3.1--9.8\% in geomean. Ablations validate the effectiveness of each individual mechanism.
\end{itemize}

\section{Background and Motivation}
\label{sec:background}

\subsection{GEMM--ReduceScatter Execution}
\label{sec:bg:tiled}
\label{sec:bg:dependence}

We use a row-parallel GEMM followed by ReduceScatter to illustrate the
dependencies between computation and
communication~\cite{coconet,decomposition-asplos23}. With $W$ GPUs, each
rank holds a slice of the GEMM reduction dimension and computes a partial
contribution to the output. ReduceScatter sums these contributions and
assigns a distinct row partition of the result to each
rank~\cite{sequence-parallel,mscclang}.
Each rank therefore produces contributions both to its
own output partition, which are consumed locally, and to other
partitions, which require inter-GPU transfer.

We call the computation that generates output tiles the \emph{producer}
and the communication that processes them the \emph{consumer}.
The producer kernel executes
as a grid of independently scheduled thread
blocks~\cite{cuda-programming-guide}. A \emph{tile} is an independently
processable output region. In the producers studied here, each tile is
computed by one block, while a block may cover multiple tiles.
Output becomes available incrementally as
blocks finish, each completing only its rank's contributions to the
tiles it covers. Reducing an output tile requires the
corresponding contributions from all ranks, but has no data dependence
on unrelated output tiles.

\subsection{Motivation}
\label{sec:bg:motivation}

In sequential execution, ReduceScatter starts only after every producer
block has completed. Across the workloads in our evaluation suite
(\S\ref{sec:eval:setup}), this organization leaves communication exposed
for \exposedShareRange{} of end-to-end latency, with a median of
\exposedShareMedian{}.

\motivationinsight{1}{Sustaining the Supply of Com\-mu\-ni\-ca\-tion-Ready Tiles}{%
Effective overlap requires computation to continuously produce
communication-ready tiles and communication to process them promptly.}

Computation progress does not always supply new data for inter-GPU
transfer. Under a traversal that groups tiles by output partition, a
producer computes a contiguous run of tiles for its own partition
(Figure~\ref{fig:intro-abstraction}, top). During this run, the GEMM
continues to make progress, but the completed tiles are consumed locally
rather than transferred to peer GPUs. This creates a gap in the
production of communication data on that GPU, reducing the portion of
its remaining computation that can overlap with inter-GPU transfer.

Waiting for unrelated output can limit the overlap gained from earlier
tile production. The communication unit need not match the producer's tile.
FlashOverlap, for example, groups one or more execution waves and
triggers a library collective after the group
completes~\cite{flashoverlap}. Grouping forms larger messages for better
bandwidth utilization. However, a tile whose required contributions are
available may still wait for unrelated output in the group. This waiting
shortens the computation window available for overlapping the tile's
communication. Tile-level signals alone cannot eliminate this wait;
communication must also process tiles independently.

\motivationinsight{2}{Balancing Overlap against Computation Slowdown}{%
Earlier communication improves overall performance only when its
benefit outweighs the additional cost imposed on computation.}
\label{sec:bg}

Creating and exploiting overlap opportunities can also reduce computational
efficiency. Changing the production order may reduce data reuse
(\S\ref{sec:evaluation}), while concurrent transfers and reductions
consume SM resources and memory bandwidth also used by
computation~\cite{gpu-maestro,resource-aware-overlap}.
Both effects can slow the production of subsequent tiles.

\begin{figure}[t!]
  \centering
  \includegraphics[width=\columnwidth]{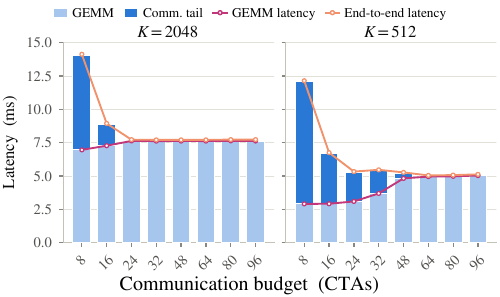}
  \caption{Effect of the communication budget on end-to-end \gemmrs{}
  latency for a longer ($K=2048$, left) and a shorter ($K=512$, right)
  computation window. Bars separate
  GEMM time from the exposed communication tail; curves report GEMM time
  and end-to-end latency. Within each panel, the tile granularity,
  producer order, and data path are fixed.}
  \Description{The two panels contain stacked bars and overlaid curves. In both
  panels the communication tail shrinks as the budget grows while GEMM
  time rises. At K equal to 2048 the end-to-end latency falls quickly and
  stays in a wide flat region; at K equal to 512 it reaches an interior
  minimum and then stops improving.}
  \label{fig:motivation}
\end{figure}

To examine the resource tradeoff, we vary a communication
\emph{budget}, defined as a cap on the number of concurrently admitted
communication blocks. Figure~\ref{fig:motivation} shows the response of
two \gemmrs{} instances with the same output shape but different
reduction dimensions $K$, yielding a longer and a shorter computation
window. Within each panel, the production order, tile granularity, and
data path are fixed, isolating the effect of communication concurrency.

With the longer window ($K=2048$), increasing the budget initially
shortens the communication tail. The reduction in the tail outweighs
the increase in GEMM time, so end-to-end latency falls before reaching
a broad near-optimal region. With the shorter window
($K=512$), increasing the budget helps only until the added GEMM delay
offsets the reduction in the communication tail. End-to-end latency
reaches an interior minimum, beyond which additional communication
capacity provides no benefit. The appropriate allocation is therefore
determined by end-to-end latency, not by the shortest
communication tail.

Section~\ref{sec:design} presents how \systemname{} coordinates these
choices to reduce end-to-end latency.

\section{Design}
\label{sec:design}

\begin{figure*}[t]
\centering
\includegraphics[width=\textwidth]{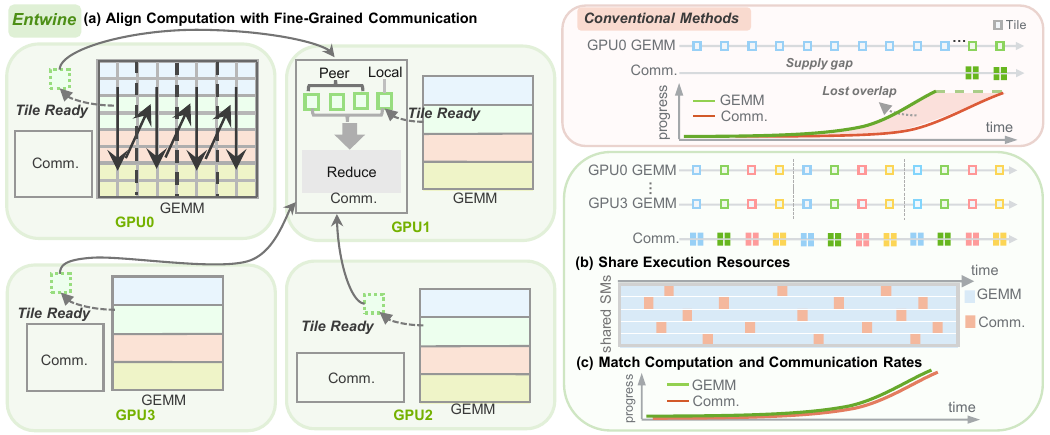}
\caption{Overview of \systemname{} for GEMM--ReduceScatter on four GPUs.
(a)~Four colors denote output partitions. For each producer, one
partition is retained locally; interleaving distributes its tiles among
those requiring remote transfer. Arrows trace peer contributions to
GPU~1's reduction.
(b)~Computation (blue) and communication (orange) blocks share an SM pool;
retiring blocks release capacity for pending work from either kernel.
(c)~Cumulative computation and communication progress under coordinated
execution. The upper-right conventional comparison marks an interval
of \emph{lost overlap}.}
\Description{Panel (a) shows four colored output partitions corresponding
to four destination GPUs. Entwine advances through these regions in an
interleaved order rather than completing one region before moving to the
next. Arrows trace contributions from peer GPUs to the reduction on
GPU 1. Panel (b) shows blue computation blocks and
orange communication blocks sharing SM capacity over time. As blocks
retire, pending work from either kernel can use the released capacity.
Panel (c) shows cumulative computation and communication progress.
The conventional comparison above it marks an interval of lost overlap.}
\label{fig:overview}
\end{figure*}

\systemname{} coordinates tiled computation and communication to minimize
end-to-end latency. Computation determines when communication can begin
for each tile. Communication shares resources with computation and can
slow the production of subsequent tiles. We therefore choose the
production order, communication granularity, and kernel configurations
together.

We present the design for GEMM--ReduceScatter. Section~\ref{sec:design:ordering}
describes the tile order and communication protocol.
Section~\ref{sec:design:sharing} explains how we select kernel
configurations under resource sharing. Section~\ref{sec:design:matching}
analyzes the communication bandwidth required at a given configuration.
Figure~\ref{fig:overview} summarizes the design, and
Figure~\ref{fig:execution-workflow} illustrates its execution.
\subsection{Coordinating Tiled Computation and Communication}
\label{sec:design:ordering}

All ranks compute partial outputs with the same dimensions. The final output
is partitioned across destination ranks. To form output tile $O[d,t]$,
destination rank~$d$ reduces the corresponding contributions $V[r,d,t]$
from all source ranks~$r$.

\begin{figure*}[t]
  \centering
  \includegraphics[width=\textwidth]{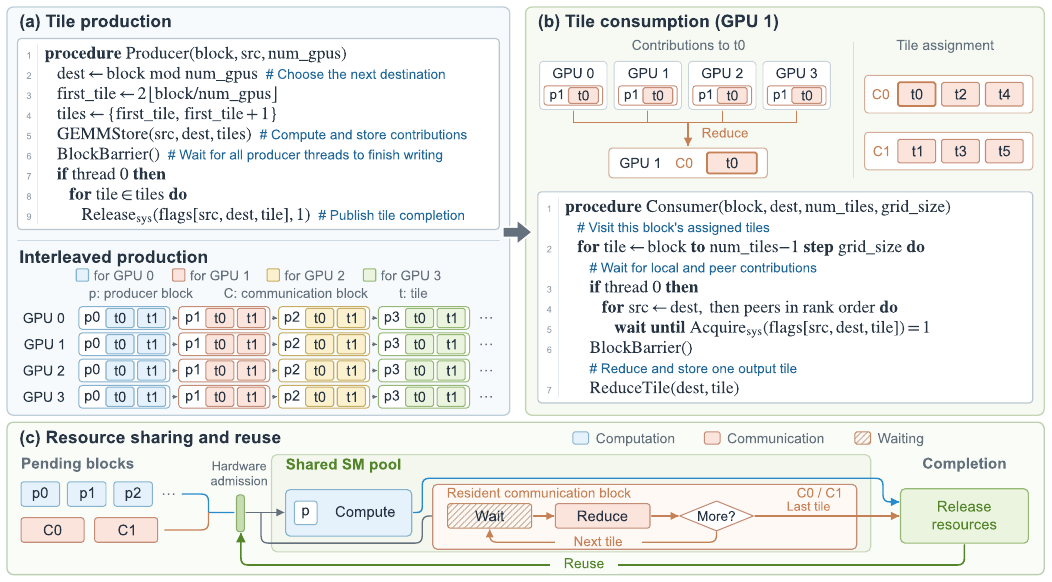}
  \caption{Tile-level execution in
  \systemname{}. The example uses four GPUs, two tiles per producer block,
  and two communication blocks per destination.
  (a)~Producer procedure and interleaved production.
  (b)~Tile consumption, dependencies, and tile assignment; the outlined $t_0$
  is expanded at upper left.
  Each communication block processes its assigned tiles in increasing
  tile-index order.
  (c)~Block lifecycles and resource reuse.
  The arrow from (a) to (b) denotes per-tile dependencies.
  Tile colors denote destinations in (a,b); colors in (c) denote kernel roles.}
  \Description{Panel (a) gives the producer mapping and publication
  protocol for a block covering two output tiles. Its outgoing arrow
  denotes per-tile dependencies on the consumer side. The lower part
  of panel (a) shows the
  first four logical producer blocks on each of four GPUs. Each block
  visits a different destination partition. Panel (b) expands only the
  dependencies of tile 0. Four source GPUs each supply a contribution
  from producer block p1 to communication block C0 on GPU 1.
  The separate task groups highlight tile 0 to connect this detail with
  its assignment. Communication block C0
  processes tiles 0, 2, and 4; C1 processes tiles 1, 3, and 5. Each tile
  is reduced after the contributions from all ranks have been published.
  Below these diagrams, the Consumer procedure waits for each tile's
  contributions and reduces it.
  Panel (c) shows hardware admission and block lifecycles within a shared
  SM resource pool. Each communication block retains its resources
  throughout its wait and reduction cycle. Completing computation and
  communication blocks release resources for pending work. The diagram
  shows lifecycle transitions rather than SM assignments or durations.}
  \label{fig:execution-workflow}
\end{figure*}

\paragraph{Interleaved production.}
With a contiguous traversal, a producer can spend an extended interval
computing only its locally retained partition, supplying no new data
for inter-GPU transfer.
\systemname{} cycles through the output partitions in logical block
order, as illustrated in Figure~\ref{fig:execution-workflow}(a).
Only one partition is local to each source rank. In the logical block
order, blocks producing local contributions are interspersed with
blocks producing contributions for peers.

For $W$ equal, block-aligned output partitions, let $p$ denote a producer
block's logical position, numbered from zero. Its destination rank $d$
and within-partition block index $\ell$ are
\begin{equation}
d = p \bmod W, \qquad \ell = \lfloor p/W \rfloor.
\label{eq:mapping}
\end{equation}
On rank~$r$, blocks with $d=r$ produce contributions to the local
partition; the others produce contributions for peers.
All ranks use the same mapping, placing contributions to the same tile
at corresponding logical positions. The destination's task order can
therefore follow a common production order across ranks.
The mapping specifies logical block order;
physical completion remains subject to GPU scheduling
(\S\ref{sec:bg:tiled}).

\paragraph{Independent tile communication.}
Each communication task reduces one output tile. The task depends on
one producer block per rank, which supplies that rank's contribution
(\S\ref{sec:eval:granularity}).
It need not wait for the rest of the destination partition.
The upper-left part of Figure~\ref{fig:execution-workflow}(b) illustrates this
dependency for tile $t_0$ in GPU~1's partition. Each rank's block $p=1$
supplies a contribution to that tile. GPU~1's communication block $C_0$
reduces the four contributions, including the one produced locally.

The producer completes its output stores before publishing a completion
signal $F[r,d,t]$ for each contribution. Before reading a tile's inputs,
the consumer block waits for the signals from all ranks
(Figure~\ref{fig:execution-workflow}(a,b)). It then reduces the tile.
This protocol ensures that reads follow the stores of all required
contributions. Section~\ref{sec:impl:produce} describes the memory
ordering that enforces it.

A computation block groups output elements to exploit data reuse and
may cover several tiles. These tiles retain separate reduction tasks,
which can be distributed across consumer blocks even when their inputs
are produced together.
This separates communication granularity from the GEMM block shape,
allowing computation to exploit reuse across multiple tiles.

\paragraph{Matching task order.}
One communication kernel processes all tile reductions for the local
output partition. Each tile is assigned to one consumer block, and each
block executes a sequence of tasks. It waits for and reduces its current
tile before advancing to the next. Blocks advance independently, without
separate kernel launches for successive tiles.

Within each output partition, number the $Q$ tiles from zero in producer
traversal order. With $x$ consumer blocks, block $j$ processes tiles
$j,j+x,j+2x,\ldots$ up to the last index below $Q$.
Each sequence follows the producer's logical traversal order, so a
consumer checks tiles in the order of their assigned producer blocks.
It still waits for its current tile if later producer blocks finish first.

The upper-right part of Figure~\ref{fig:execution-workflow}(b) shows this
assignment with $x=2$. Block $C_0$ processes $t_0,t_2,t_4$ in sequence,
while $C_1$ processes $t_1,t_3,t_5$. In this example, each producer block
covers two tiles. On every rank, block $p=1$ produces contributions
to $t_0$ and $t_1$.
By Equation~(\ref{eq:mapping}), the next visits to GPU~1's partition use
blocks $p=5$ and $p=9$, which supply the next two pairs.
Each consumer thus takes one tile from each pair.

Changing $x$ redistributes the same tile tasks among consumer blocks.
A smaller grid gives each block a longer sequence without combining
the dependencies of successive tiles. Once its current tile's inputs
are available, a block can reduce that tile while later tiles are still
being computed.
The GEMM block shape determines which tiles are produced together;
$x$ determines how their reductions are distributed across blocks.

\paragraph{Data reuse.}
Interleaving across output partitions may cause successive producer
blocks to access different row panels of $A$, reducing opportunities for
reuse. The choice of GEMM block shape therefore affects the tradeoff
between data reuse and the overlap enabled by interleaving
(\S\ref{sec:eval:ordering}). We assess this tradeoff under concurrent
execution, where both kernels also compete for resources.

\subsection{Optimization under Resource Sharing}
\label{sec:design:sharing}

\systemname{} executes one computation kernel and one communication
kernel concurrently on each GPU. Separate kernels allow communication
concurrency to vary without changing the GEMM block shape. Their
performance remains coupled through shared SM resources and memory
bandwidth~\cite{parallelkittens,ficco}.

\paragraph{Communication budget.}
The communication budget $x$ is the grid size of the communication
kernel, capped by the number of tiles in the rank's output partition.
The thread count $b$ controls parallelism within each communication
block and affects its resource requirements. Together, $x$ and $b$
determine how much communication work can compete with GEMM.

Communication blocks retain their SM resources while waiting for inputs.
The budget must therefore leave capacity for computation to produce
those inputs (\S\ref{sec:impl:sharing}). The grid size stays constant
throughout an invocation, while the number of resident communication
blocks changes as blocks are admitted and complete. No fixed set of
SMs is reserved for communication.

\paragraph{Resource reuse.}
Blocks from both kernels share an SM pool
(Figure~\ref{fig:execution-workflow}(c)). The hardware scheduler admits
pending work from either kernel as resources become
available~\cite{cuda-programming-guide}.
A computation block releases its resources after computing its output
region. A communication block retains its resources across successive
tasks and releases them after its last reduction. For example, $C_0$
keeps its allocation while waiting for $t_2$ and releases it after
reducing $t_4$. Pending computation can then use the released capacity
while other communication blocks remain active.

\paragraph{End-to-end optimization.}
\label{sec:design:selection}
Larger budgets allow more communication blocks to make progress, but may
slow computation through competition for SM resources and memory
bandwidth (Figure~\ref{fig:motivation}). The global communication tail
is the interval between the last computation kernel finishing and all
communication completing. Increasing concurrency can shorten this
tail while extending the computation window.

The GEMM block shape also affects this tradeoff through data reuse and
the number of tiles produced together. We therefore evaluate GEMM
configurations together with the communication budget $x$ and thread
count $b$. \systemname{} selects the combination with the lowest
end-to-end latency under concurrent execution. This criterion accounts
for both communication progress and GEMM slowdown, which standalone
kernel measurements do not capture
(\S\ref{sec:eval:coupling}, \S\ref{sec:eval:envelope}).
Section~\ref{sec:impl:sharing} describes the offline profiling procedure.

\subsection{Communication Bandwidth Analysis}
\label{sec:design:matching}

We now relate achieved GEMM throughput to the average communication
bandwidth required during GEMM execution. This analysis explains the
bandwidth requirements of configurations selected by end-to-end latency
(\S\ref{sec:design:selection}).

Consider one rank computing an $M\times N$ partial output over local
reduction dimension $K$.
Finalizing its $1/W$ output partition reads contributions from $W-1$
peers, with total remote-read volume
\begin{equation}
  V_{\mathrm{comm}}=\frac{W-1}{W}MNs,
  \label{eq:vcomm}
\end{equation}
where $s$ is the size of a stored contribution element in bytes.
Let $\Phi(x,K)$ denote this rank's achieved GEMM throughput during
concurrent execution, measured in FLOP/s. It includes the slowdown caused
by communication. Its dependence on output shape, datatype, rank count,
GEMM configuration, and communication thread count is implicit.
The rank's GEMM latency is
\begin{equation}
  T_G(x,K)=\frac{2MNK}{\Phi(x,K)}.
  \label{eq:tg}
\end{equation}
The average bandwidth required to process the remote-read volume within
this window is
\begin{equation}
  \Breq(x,K)=\frac{V_{\mathrm{comm}}}{T_G(x,K)}
  =\frac{W-1}{W}\frac{s\Phi(x,K)}{2K}.
  \label{eq:breq}
\end{equation}
At fixed output dimensions, rank count, and datatype, increasing $K$
adds computation without adding communication volume. The required
bandwidth falls when GEMM throughput grows more slowly than $K$.
Conversely, a shorter computation window requires higher average
bandwidth for the same volume. The explicit factors $M$ and $N$ cancel,
but output shape still affects the required bandwidth through GEMM
throughput.

The budget $x$ also affects the required bandwidth through GEMM throughput.
If additional communication blocks slow GEMM, the computation window
lengthens and the same volume requires a lower average bandwidth.
Lower required bandwidth can therefore reflect slower GEMM, even when
the communication volume is unchanged. This is why configuration
selection considers end-to-end latency.

Equation~(\ref{eq:breq}) averages over the full computation window.
Different production orders have the same required average bandwidth
when communication volume and GEMM latency are unchanged. However, each
tile can be reduced only after its inputs become available. Late inputs
leave less computation to overlap with that reduction. The tile order
in \S\ref{sec:design:ordering} therefore matters even when the average
bandwidth requirement is unchanged. Section~\ref{sec:eval:sizing}
compares required and measured communication bandwidth, and
\S\ref{sec:eval:coupling} examines GEMM slowdown as the budget increases.

\section{Implementation}
\label{sec:implementation}

\begin{figure*}[t]
  \centering
  \includegraphics[width=\textwidth]{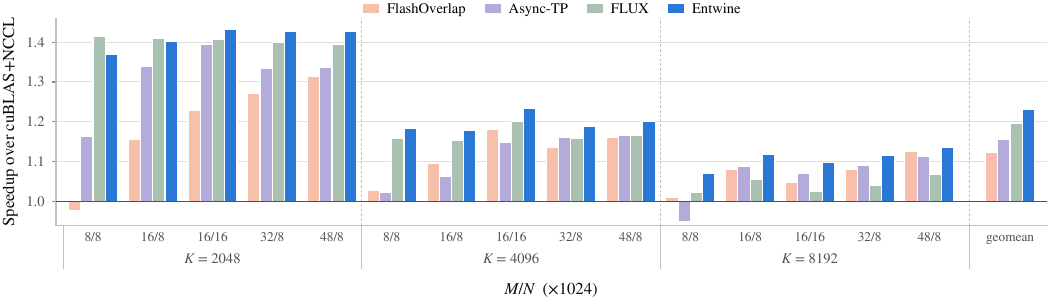}
  \caption{End-to-end \gemmrs{} speedup over cuBLAS~+~NCCL.
  Workloads are grouped by local GEMM reduction dimension $K$;
  the rightmost group reports the geomean.}
  \Description{Grouped bars compare Entwine, FlashOverlap, Async-TP, and FLUX
  across the 8-GPU operator suite and its overall geomean.}
  \label{fig:operator-performance}
\end{figure*}

\begin{figure}[t]
  \centering
  \includegraphics[width=\columnwidth]{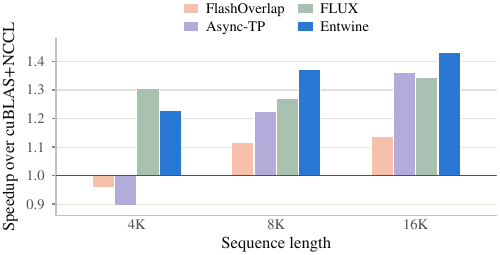}
  \caption{Layer-level speedup over cuBLAS~+~NCCL for the Llama 3
  70B attention-output GEMM, including residual addition and RMSNorm.
  Sequence lengths are global, before tensor-parallel partitioning.}
  \Description{Grouped bars compare Entwine, FlashOverlap, Async-TP, and FLUX at
  sequence lengths 4K, 8K, and 16K. Entwine trails FLUX at 4K and leads at 8K and
  16K.}
  \label{fig:layer-performance}
\end{figure}

\subsection{Execution Path}
\label{sec:impl:overview}

The prototype pairs a CUTLASS~\cite{cutlass} GEMM with a CUDA reduction kernel
for row-parallel \gemmrs{} (Figure~\ref{fig:execution-workflow}).
It targets NVIDIA SM80 GPUs within a single peer-accessible NVLink domain.
Output partitions have equal row counts. Each partition's row count and
the output width are multiples of 128.

Each rank allocates its partial output and publication flags in a
symmetric CUDA IPC heap. Contributions and flags occupy matching offsets
in the heaps on all ranks. The reducer uses these offsets to read
contributions directly from peer memory. Allocations are reused across
invocations. Before each invocation, ranks reset their flags and
synchronize.

Each invocation launches the GEMM and reduction kernels on independent CUDA
streams after a common start event. We capture their kernel launches in CUDA
Graphs~\cite{cuda-programming-guide} to reduce repeated host-launch overhead.

\subsection{Producer Traversal, Publication, and Reduction}
\label{sec:impl:produce}

Figure~\ref{fig:execution-workflow} presents the \emph{Producer} and
\emph{Consumer} procedures, each executed cooperatively by one thread
block on its source or destination rank. Their \emph{block} arguments
denote producer position $p$ and consumer index $j$, respectively.
The parameters \emph{num\_gpus}, \emph{num\_tiles}, and \emph{grid\_size}
denote $W$, $Q$, and $x$.
The producer retains the selected GEMM kernel's tensor-core main loop.
Figure~\ref{fig:execution-workflow}(a) uses \emph{GEMMStore} to denote
computing and storing a producer block's contributions.
A custom CUTLASS thread-block swizzle implements the traversal in
Equation~(\ref{eq:mapping}) and Figure~\ref{fig:execution-workflow}(a).

The reduction unit is a $128\!\times\!128$ output tile.
Producer blocks cover either $128\!\times\!128$ or
$128\!\times\!256$ outputs. These configurations differ in data reuse
under the interleaved order (\S\ref{sec:eval:ordering}). An epilogue
visitor~\cite{evt} publishes completion signals. The wider block
publishes both tile flags after completing its output stores, as in
Figure~\ref{fig:execution-workflow}(a). The consumer uses the same tile
size with either producer.

Producer threads synchronize after their output stores, so publication
covers the writes of the entire thread block (CTA).
Thread~0 then publishes the tile flags with system-scope release stores.
The pseudocode writes $F[r,d,t]$ as \emph{flags[src,dest,tile]} and denotes
system scope as \emph{sys}.

For each tile in its assigned sequence, consumer thread~0 polls the
local flag and then the peer flags using system-scope acquire loads.
The local flag is necessary because the producer and consumer execute
asynchronously on the same GPU.

After observing all $W$ flags for the current tile, thread~0 synchronizes
with its CTA, extending the acquire ordering to all threads that read
the tile.
The release stores, acquire loads, and CTA barriers thus order these
reads after the corresponding producer stores~\cite{cuda-programming-guide,ptx-memory-model}.

The CTA cooperatively reduces the tile through \emph{ReduceTile}
in Figure~\ref{fig:execution-workflow}(b).
Each thread processes a disjoint subset of tile elements. For each
element, it initializes an accumulator from the local contribution,
adds peer values in rank order, and stores the result in the local output
partition. Each output element therefore has a single writer.
The implementation distributes packed FP16 pairs across threads in a
strided loop. For two ranks, it groups these pairs into vector loads and
stores.

\begin{figure*}[t]
  \centering
  \includegraphics[width=\textwidth]{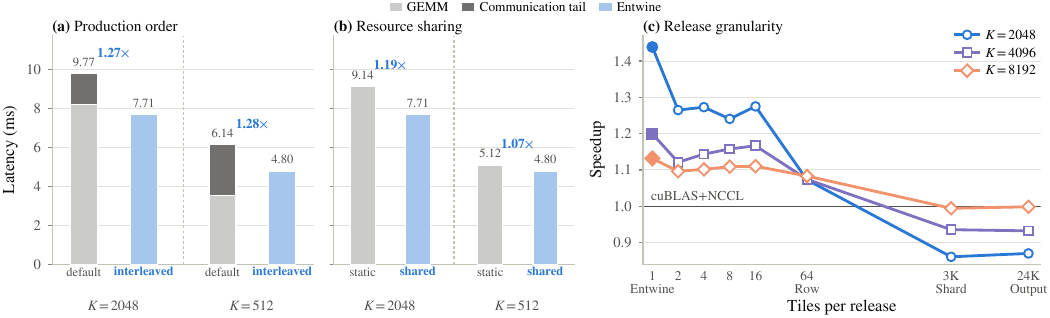}
  \caption{Effects of computation order, resource sharing, and release
  granularity. Panels (a) and (b) show end-to-end latency with the
  communication tail highlighted. Panel (c)
  reports end-to-end speedup as the release unit grows from 1 tile to the
  whole output, with a fixed one-tile reduction task. Each result includes
  the synchronization required by its release policy.}
  \Description{Panel (a) compares the default and interleaved computation
  orders at 2 values of $K$; interleaving lowers the communication tail
  and total time. Panel (b) compares static
  partitions and shared resource pools for the same workloads as panel (a);
  the shared pool lowers total time in both cases. Panel (c) shows 3 curves
  in which a single tile gives the highest speedup for every workload, while full output
  partitions and the whole output approach or fall below the non-overlapped baseline.}
  \label{fig:design}
\end{figure*}

\subsection{Kernel Launch and Configuration}
\label{sec:impl:sharing}

The GEMM kernel runs on a default-priority stream and the reduction kernel
on a high-priority stream. Stream priority favors pending reducer work as SM
capacity becomes available (Figure~\ref{fig:execution-workflow}(c)).
The hardware scheduler determines block admission, and running blocks
continue to completion~\cite{cuda-programming-guide}.
The default reducer grid is capped below the device's SM count.
This leaves capacity for GEMM blocks without reserving a fixed set
of SMs, even while reducer CTAs wait for inputs.

For each matrix shape and rank count, offline profiling evaluates
GEMM kernels and reducer grid sizes with 128, 256, or 512 threads
per reducer block. We select the combination with the lowest end-to-end latency
under concurrent execution (\S\ref{sec:design:selection}).

\section{Evaluation}
\label{sec:evaluation}

We evaluate whether \systemname{}'s coordination of tiled computation and
fine-grained communication reduces end-to-end latency. We compare against a
sequential reference and three existing overlap approaches at both operator
and layer levels. Controlled ablations examine how computation order, release
granularity, and resource sharing affect this benefit, and budget sweeps
characterize the tradeoff between communication progress and computation slowdown.

\subsection{Experimental Setup}
\label{sec:eval:setup}

\paragraph{Testbed.}
Experiments use 2, 4, or 8 NVIDIA \gpuname{} GPUs on one NVLink-connected
node, with 8 as the default. We use CUDA~\cudaver{}, cuBLAS~\cublasver{},
NCCL~\ncclver{}, CUTLASS at commit \cutlassver{}, FlashOverlap at commit
\flashoverlapcommit{}, Async-TP from PyTorch~\torchver{}~\cite{pytorch}, and FLUX at commit
\fluxcommit{}.

\paragraph{Workloads.}
Two workload suites cover different ratios of computation to communication
and different LLM GEMM shapes
(Table~\ref{tab:workloads}). The main suite uses the
MLP output widths of Llama 3 70B and Llama 3.1 405B~\cite{llama3} and covers
15 workloads spanning a $6\times$ range in output size $MN$ and a $4\times$
range in per-rank reduction dimension $K$.
At a fixed tensor-parallel width and datatype, increasing $MN$ scales both
GEMM work and ReduceScatter volume, whereas increasing $K$ adds computation
without adding communication. Equal-volume outputs with different aspect
ratios separate output geometry from the amount of work.

The model-derived suite includes the attention-output and MLP-output GEMMs of
Llama 3 8B and 70B~\cite{llama3} and the MLP-output GEMM of
Qwen2.5-72B~\cite{qwen25}. Five logical activation sizes per GEMM yield
25 workloads. These cover shorter computation windows, complementing the
main suite in the operating-range analysis (\S\ref{sec:eval:envelope}).

\begin{table}[t]
  \caption{Evaluation workloads on 8 GPUs. $K$ is the per-rank reduction
  dimension; all methods pad model-derived cases with logical $M=512$ to
  physical $M=1024$.}
  \label{tab:workloads}
  \Description{The main operator suite uses the MLP output widths of Llama 3
  70B and Llama 3.1 405B. The model-derived suite uses attention-output and
  MLP-output GEMMs from Llama 3 8B and 70B and Qwen2.5-72B. Model-derived
  512-row cases are padded to 1024 rows for all methods.}
  \footnotesize
  \begin{tabular}{@{}l@{\hspace{4pt}}l@{}}
    \toprule
    \multicolumn{2}{@{}l}{\textbf{Main operator suite} --- 15 workloads} \\
    Llama 3 70B width, $N=8192$ & $M = 8192$,\, $16384$,\,
      $32768$,\, $49152$ \\
    Llama 3.1 405B width, $N=16384$ & $M = 16384$ \\
    Per-rank reduction $K$ & $2048$,\, $4096$,\, $8192$ \\
    \midrule
    \multicolumn{2}{@{}l}{\textbf{Model-derived suite at TP=8} --- 25 workloads}
      \\
    GEMM & $(N,K)$ \\
    Llama 3 8B attn. output & $(4096,\,512)$ \\
    Llama 3 8B MLP output     & $(4096,\,1792)$ \\
    Llama 3 70B attn. output & $(8192,\,1024)$ \\
    Llama 3 70B MLP output    & $(8192,\,3584)$ \\
    Qwen2.5-72B MLP output    & $(8192,\,3696)$ \\
    \addlinespace
    Envelope sweep, logical $M$ & 512,\, 1024,\, 2048,\, 4096,\,
      8192 \\
    \bottomrule
  \end{tabular}
\end{table}

\paragraph{Baselines.}
The sequential reference, \emph{cuBLAS + NCCL}, completes the GEMM before
invoking NCCL ReduceScatter~\cite{cublas,nccl}. \emph{FlashOverlap} starts
an asynchronous collective after releasing each completed wave
group~\cite{flashoverlap}. \emph{Async-TP} pipelines ReduceScatter behind
chunked GEMM execution through PyTorch's symmetric-memory
operator~\cite{asynctp}. \emph{FLUX} integrates reduction into the GEMM epilogue and
autotunes tiling~\cite{flux}. We use each method's native implementation
and recommended workload-specific tuning procedure.

\paragraph{Protocol.}
Unless stated otherwise, speedups are relative to cuBLAS~+~NCCL on the same
workload, and geomeans weight workloads equally.
We time \iterscount{} iterations with CUDA events after \warmupcount{}
warmups and report the maximum median latency across ranks. We validate
outputs against an unfused reference with absolute tolerance \corrAtol{}
and relative tolerance \corrRtol{}.

\begin{figure*}[t]
  \begin{minipage}[t]{\columnwidth}
  \centering
  \includegraphics[width=\columnwidth]{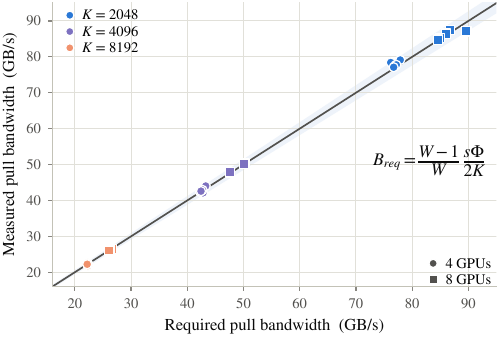}
  \caption{Measured communication bandwidth versus average demand computed from the
  measured computation window (Equation~(\ref{eq:breq})).
  Colors denote local $K$; markers denote tensor-parallel
  width. The diagonal indicates equal measured and required bandwidth.}
  \Description{Scatter plot of measured and required pull bandwidth for 30
  measurements. Points for 3 values of $K$ and 2 tensor-parallel widths lie
  close to the diagonal, with the 5 output shapes in each group clustered
  together.}
  \label{fig:demand-validation}
  \end{minipage}\hfill
  \begin{minipage}[t]{\columnwidth}
  \centering
  \includegraphics[width=\linewidth]{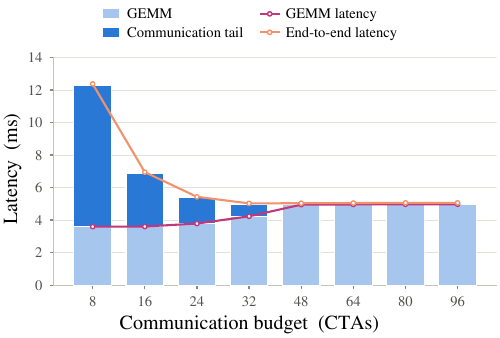}
  \caption{Effect of communication budget on latency at $K=768$. Bars show computation
  windows and communication tails; curves show GEMM and end-to-end latency,
  including completion overhead. From 32 to 48 CTAs, computation slowdown offsets
  the shorter tail.}
  \Description{At K equal to 768, increasing the communication budget from
  32 to 48 CTAs nearly removes the communication tail, but a longer
  computation window leaves end-to-end latency almost unchanged.}
  \label{fig:resource-coupling}
  \end{minipage}
\end{figure*}

\subsection{Overall Performance}
\label{sec:eval:performance}

\subsubsection{Operator-Level Performance}
\label{sec:eval:operator}

Across the 15 main-suite workloads in Figure~\ref{fig:operator-performance},
\systemname{} achieves a \tpeightOursSpeedup{} geomean speedup over
cuBLAS~+~NCCL and a \tpeightOracleRatio{} geomean speedup over the
per-workload best among prior methods. It outperforms all three overlap
approaches on \tpeightOursWins{} workloads. Geomean latency decreases
relative to FlashOverlap, Async-TP, and FLUX by \tpeightFOLatencyDelta{},
\tpeightAsyncLatencyDelta{}, and \tpeightFLUXLatencyDelta{}, respectively.

Across the suite,
overlap efficiency---the ratio of uncontended GEMM time to end-to-end
latency---reaches \tpeightOverlapEfficiency{} in geomean, indicating that
\systemname{} overlaps GEMM and ReduceScatter while keeping their combined
latency close to that of standalone GEMM execution.

Speedup over sequential execution decreases as per-rank $K$ increases at
fixed output dimensions. Larger $K$ extends the computation window without
adding communication volume, so communication accounts for a smaller share
of sequential execution. This reduces the speedup available from overlap.
Output size also affects the comparison with FLUX: at $K=2048$, FLUX
outperforms \systemname{} on the two smallest outputs, which provide less
computation to overlap reduction. \S\ref{sec:eval:envelope} examines
performance across a wider range of workloads.

\subsubsection{Layer-Level Performance}
\label{sec:eval:layer}

To assess whether the overlap benefits carry over to a longer operator
chain, we evaluate the Llama 3 70B attention output, including the GEMM and
ReduceScatter followed by residual addition and RMSNorm. We vary sequence
length while keeping the model and GEMM output width fixed
(Figure~\ref{fig:layer-performance}). \systemname{} reduces latency
relative to cuBLAS~+~NCCL, FlashOverlap, and Async-TP at all three sequence
lengths.
Relative to FLUX, \systemname{} lowers layer-level latency by
\layerEightKFluxGain{} and \layerSixteenKFluxGain{} at
8K and 16K, respectively, whereas FLUX has \layerFourKFluxLead{} lower
latency at 4K. Longer sequences increase the GEMM wave count, providing a wider
computation window over which reductions can progress concurrently.
These results demonstrate that the benefits of \systemname{}'s fine-grained
overlap extend to layer-level execution.

\begin{figure*}[t]
  \begin{minipage}[t]{\columnwidth}
  \centering
  \includegraphics[width=\columnwidth]{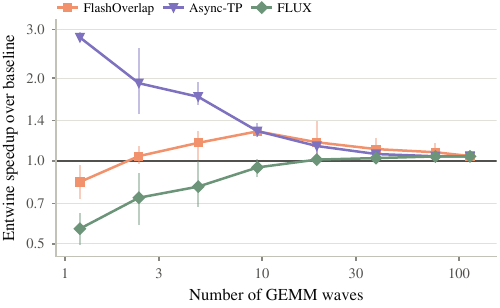}
  \caption{Speedup versus GEMM wave count. Marks show \systemname{}'s geomean speedup over each
  baseline at the same GEMM wave count; whiskers show the full workload range.}
  \Description{Entwine speedup over FlashOverlap, Async-TP,
  and FLUX versus the number of GEMM waves, across the 25 model-derived
  workloads and the 15 workloads in the main operator suite. Entwine's
  geomean speedup crosses 1 at lower wave counts against FlashOverlap
  and at higher wave counts against FLUX.}
  \label{fig:operating-range}
  \end{minipage}\hfill
  \begin{minipage}[t]{\columnwidth}
  \centering
  \includegraphics[width=\linewidth]{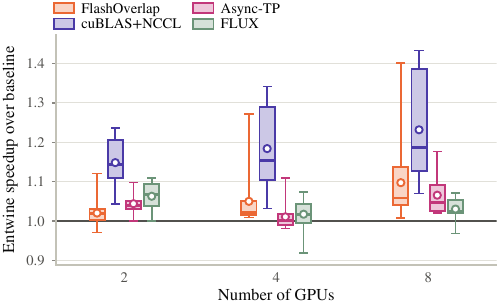}
  \caption{\systemname{} speedup across tensor-parallel widths,
  with per-rank $M$, $N$, and $K$ fixed. Boxes show the
  interquartile range, lines the medians, whiskers the full range, and
  hollow circles the geomeans.}
  \Description{Box plots of Entwine speedup over cuBLAS plus NCCL,
  FlashOverlap, Async-TP, and FLUX for the same 15 main-suite workloads at 2,
  4, and 8 GPUs.}
  \label{fig:tp-scaling}
  \end{minipage}
\end{figure*}

\subsection{Mechanism Ablations}
\label{sec:eval:analysis}

The ablations in Figure~\ref{fig:design} isolate computation order, release
granularity, and resource sharing. Each comparison changes one policy while
holding the workload, tiling, data path, and consumer task assignment fixed.
For policies with tunable communication concurrency, we sweep the same budget
range and report each policy's lowest end-to-end latency.

\subsubsection{Computation Order}
\label{sec:eval:ordering}

We compare \systemname{}'s interleaved mapping in Figure~\ref{fig:design}(a),
which follows the consumer's $N$-major order within each partition,
with the producer kernel's default $M$-major traversal. By coordinating tile
production with reduction, \systemname{} lowers end-to-end latency in both
workloads, including the case in which GEMM execution slows.
At $K=\tileOrderShortK$, the GEMM takes longer under interleaving, but
the tail shrinks by \tileOrderShortTailReduction{}, yielding a net
\tileOrderShortSpeedup{} end-to-end improvement.
At $K=\tileOrderLongK$, both the computation window and the tail shrink,
improving end-to-end performance by \tileOrderLongSpeedup{} with a
\tileOrderLongTailReduction{} tail reduction. The default order leaves
a substantial tail even at its best budget.

\subsubsection{Release Granularity}
\label{sec:eval:granularity}
To examine how release granularity affects overlap, we vary the number of
tiles released together while keeping each reduction task fixed at one tile
(Figure~\ref{fig:design}(c)). A group is released when its last tile
finishes. This comparison changes when reductions can begin while
preserving the amount of work in each reduction task.

Tile-level release gives the highest speedup on all 3 workloads:
\granularityKtwoSpeedup{}, \granularityKfourSpeedup{}, and
\granularityKeightSpeedup{} over the non-overlapped baseline. Small groups
retain much of this benefit, whereas releasing a full output partition or the
whole output brings performance to or below the baseline. Larger groups
delay reductions on tiles completed earlier in the group, leaving less
computation to overlap when those reductions begin. With the reduction tasks
held fixed, \systemname{}'s tile-level release preserves overlap that is lost
under partition- or output-level release.

\subsubsection{SM Resource Sharing}
\label{sec:eval:partition}

Figure~\ref{fig:design}(b) compares shared-SM execution with static
partitioning on the same two workloads as the computation-order ablation.
Both policies use the same GEMM and reduction kernels, task order, and data path.
We select the best communication budget for each policy independently.
Static partitioning reserves a fixed subset of SMs for communication.
Reduction blocks remain resident until GEMM finishes, even after completing
their assigned reductions.\footnote{We enqueue reduction blocks before GEMM
and allocate enough dynamic shared memory to allow exactly one reduction
block per occupied SM.}
Shared-SM execution removes these restrictions and lets both kernels use
the same SM pool.

At their selected budgets, both policies leave little communication tail.
Shared-SM execution has a shorter computation window, improving end-to-end
performance by
\sharedPartitionLongSpeedup{} at $K=\tileOrderLongK$ and
\sharedPartitionShortSpeedup{} at $K=\tileOrderShortK$.
Under shared-SM execution, SMs become available to pending GEMM blocks
after reductions finish. This allows \systemname{} to sustain communication
progress with less GEMM slowdown than static partitioning.

\subsection{Communication Bandwidth and Concurrency}
\label{sec:eval:matching}

\subsubsection{Matching Communication to Computation}
\label{sec:eval:sizing}

At the selected budgets, measured communication bandwidth closely matches
the average communication demand. We compute this demand from each rank's
measured computation window using Equation~(\ref{eq:breq}).
Measured bandwidth is the remote-read volume divided by the communication
kernel's elapsed time, including waits for tile inputs.
Across the 15 main-suite
workloads at both 4 and 8 GPUs, required and measured bandwidth differ by at
most \modelPointErr{}, with a median gap of \modelMedianErr{}
(Figure~\ref{fig:demand-validation}). This agreement indicates that
\systemname{}'s selected configurations sustain communication at
approximately the rate needed to process output during the computation
window.

Within each $(W,K)$ group, the 5 output shapes achieve similar GEMM
throughput and cluster around a common bandwidth demand despite their
different communication volumes. This clustering is consistent with
Equation~(\ref{eq:breq}), in which the output dimensions cancel and shape
affects demand through achieved GEMM throughput. At comparable
throughput, larger $K$ lowers the required rate by spreading communication
over a longer computation window.

\subsubsection{Effect of Communication Concurrency}
\label{sec:eval:coupling}

A short\-er communication tail does not necessarily reduce end-to-end
latency. At $K=768$ (Figure~\ref{fig:resource-coupling}), increasing the
budget from 32 to 48 CTAs nearly eliminates the tail, but the accompanying
computation slowdown offsets this gain, leaving total latency almost unchanged.
The full budget sweep across eight values of $K$
(Figure~\ref{fig:resource-coupling-full}) shows the same pattern more
broadly. For larger $K$, a larger budget initially shortens the
tail with little effect on computation, and end-to-end latency reaches a
broad plateau. For smaller $K$, more concurrent consumer work
increasingly delays the producer and the output on which later reductions
depend.

With all inputs available and no competing producer, communication bandwidth rises
with the budget before reaching a plateau (Figure~\ref{fig:isolated-response}).
Beyond that plateau, additional concurrency provides little communication
capacity, while concurrent consumer work can still delay the producer through
resource occupancy and memory traffic. Isolated bandwidth therefore does
not capture the end-to-end effect of added communication concurrency.
This supports selecting computation and communication configurations by
end-to-end latency (\S\ref{sec:design:selection}), which reflects both
communication progress and computation slowdown.

\subsection{Sensitivity Analysis}
\label{sec:eval:sensitivity}

Workload size and tensor-parallel width change the computation available for
overlap and the communication that must complete within it.

\subsubsection{Operating Range}
\label{sec:eval:envelope}

Figure~\ref{fig:operating-range} combines both workload suites to characterize
how \systemname{}'s overlap benefits vary with the available producer work.
We group results by GEMM wave
count, defined as the producer's thread-block count divided by the number of SMs.
The 25 model-derived workloads span \waveProjectionSweepRange{}
waves, and the 15 main-suite workloads span \waveOperatorRange{} waves,
with the two ranges meeting at \waveSuiteMeet{} waves. In addition to this
variation in block count, $K$ affects the bandwidth required during the
computation window (\S\ref{sec:eval:sizing}).

\systemname{}'s geomean speedup over FlashOverlap crosses 1 in the
\envelopeFO{}-wave interval, and its geomean speedup over FLUX crosses 1
in the \envelopeFLUX{}-wave interval. It remains ahead of Async-TP
throughout the measured range.
With few producer waves, little computation remains after contributions
become available, limiting the time available for concurrent reductions.
As the wave count grows, \systemname{} overlaps reductions with the remaining
GEMM computation over a longer window (\S\ref{sec:eval:analysis}).
The sequence-length trend in the layer-level results is consistent with
this pattern (Figure~\ref{fig:layer-performance}).

\subsubsection{Tensor-Parallel Width}
\label{sec:eval:scaling}

Figure~\ref{fig:tp-scaling} compares 2, 4, and 8 GPUs while holding each
workload's per-rank GEMM dimensions fixed. This preserves local computation
as the collective combines contributions from more ranks, increasing the
remote data each rank must read (Equation~(\ref{eq:vcomm})).
\systemname{} outperforms all baselines in geomean at every tested width.
Its geomean speedup over FlashOverlap grows from \scalingFOtwo{} at 2 GPUs
to \scalingFOeight{} at 8 GPUs, and over cuBLAS~+~NCCL from
\scalingTorchtwo{} to \scalingTorcheight{}. Across these widths,
\systemname{} thus maintains its geomean performance advantage as
communication demand grows relative to the fixed local GEMM workload.

\section{Related Work}
\label{sec:related-work}

\noindent\textbf{Communication Granularity and Dependencies.}
Decomposing dependent operations creates opportunities for overlap:
communication on one part of a tensor can proceed while computation
continues on another.
CoCoNet jointly optimizes computation and communication through program
transformations and joint kernel generation~\cite{coconet}.
Wang et al.~\cite{decomposition-asplos23} and Domino~\cite{domino}
decompose dependent work into smaller operations that can be pipelined.
Centauri partitions communication across primitives, device groups, and
workloads~\cite{centauri}.
SYNDICATE divides communication into motifs and jointly optimizes their
scheduling and execution plans~\cite{syndicate}.
FLUX~\cite{flux} and Punniyamurthy et al.~\cite{fused-collective} fuse
computation with communication, while T3~\cite{t3} uses hardware to track
output production and trigger communication.
FlashOverlap communicates one or more completed execution waves at a time,
using output reordering to form contiguous communication
buffers~\cite{flashoverlap}.
Cui et al.~\cite{moe-tile-signaling} signal completion per tile but
transfer larger segments spanning one or more full-width row bands in MoE.
Both balance communication start time against transfer efficiency.
\systemname{} instead communicates output tiles independently, avoiding
waits for unrelated output in a larger group.
TileLink~\cite{tilelink}, Triton-Distributed~\cite{triton-distributed},
and Syncopate~\cite{syncopate} provide abstractions for expressing
dependencies and generating overlapped execution.

\noindent\textbf{Communication-Aware Computation Ordering.}
Computation order determines when communication inputs become available.
Ordering policies depend on whether computation produces data for
communication or consumes data received from other GPUs.
CoCoNet computes MatMul output chunks in the order required by the
collective~\cite{coconet}.
FLUX uses tile swizzling to reduce conflicting writes across GPUs or
align computation with incoming data~\cite{flux}.
Comet schedules computation on local inputs while remote inputs
arrive~\cite{comet}.
For MoE return communication, Cui et al. prioritize output destined for
remote GPUs~\cite{moe-tile-signaling}.
\systemname{} reorders tile computation to supply communication data at a
more regular pace, creating overlap opportunities throughout computation.
TileLink exposes tile order as a design choice, including tradeoffs between
data reuse and waiting for inputs~\cite{tilelink}.
Syncopate rewrites tile schedules to follow a communication
plan~\cite{syncopate}.

\noindent\textbf{Resource Coordination for Overlap.}
Concurrent communication competes with computation for SM resources and
memory bandwidth.
Comet assigns specialized thread blocks to computation and communication
and tunes their allocation~\cite{comet}.
Cui et al. partition SMs between two persistent kernels for
MoE~\cite{moe-tile-signaling}.
\systemname{} shares SM resources between computation and communication,
using a communication budget to balance communication progress against
computation slowdown.
ParallelKittens examines communication mechanisms and resource scheduling,
including overlap within an SM and across separate SMs~\cite{parallelkittens}.
HFUSE combines kernels with complementary resource demands through
horizontal fusion~\cite{hfuse}.
Cui and Peric\`as shape computation kernel residency and raise communication
stream priority~\cite{resource-aware-overlap}.
FiCCO characterizes decomposition overhead and resource contention, and
uses DMA offloading to improve overlap~\cite{ficco}.
ACE, ARK, and T3 reduce communication demands on compute resources through
hardware support~\cite{ace,ark,t3}, while NVSHMEM and MSCCL++ expose
lower-level transfer and synchronization
primitives~\cite{nvshmem,mscclpp}.

\section{Conclusion}
\label{sec:conclusion}

Effective overlap requires computation to produce communicable tiles at a
steady rate and communication to process them promptly. The benefit of
overlap must also outweigh the accompanying computation slowdown.
\systemname{} addresses these coupled requirements through interleaved tile
production, independent tile-level communication, and shared-SM execution
with a communication-concurrency budget. Across representative
tensor-parallel LLM workloads, \systemname{} achieves a geomean speedup of
\tpeightOursSpeedup{} (up to \tpeightVsTorchMax{}) over
cuBLAS+\allowbreak NCCL, and outperforms state-of-the-art overlap baselines
by 3.1--9.8\% in geomean.

\section*{Acknowledgments}
ChatGPT was used to assist with manuscript preparation, including
language editing and \LaTeX{} formatting. All resulting content was reviewed
and approved by the authors.

\bibliographystyle{ACM-Reference-Format}
\bibliography{bibliography/references}

@misc{llama3,
  author       = {{Meta AI}},
  title        = {{Llama 3 and Llama 3.1} Model Configurations},
  year         = {2025},
  howpublished = {Model artifact, commit 0e0b8c5},
  url          = {https://github.com/meta-llama/llama-models/blob/0e0b8c5/models/sku_list.py}
}

@misc{qwen25,
  author       = {{Qwen Team}},
  title        = {{Qwen2.5-72B} Model Configuration},
  year         = {2024},
  howpublished = {Model artifact, revision efba10c},
  url          = {https://huggingface.co/Qwen/Qwen2.5-72B/blob/efba10c/config.json}
}

@inproceedings{mesh-tensorflow,
  author    = {Noam Shazeer and Youlong Cheng and Niki Parmar and Dustin Tran and
               Ashish Vaswani and Penporn Koanantakool and Peter Hawkins and
               HyoukJoong Lee and Mingsheng Hong and Cliff Young and
               Ryan Sepassi and Blake Hechtman},
  title     = {{Mesh-TensorFlow}: Deep Learning for Supercomputers},
  booktitle = {Advances in Neural Information Processing Systems},
  volume    = {31},
  publisher = {Curran Associates, Inc.},
  year      = {2018},
  url       = {https://proceedings.neurips.cc/paper/2018/hash/3a37abdeefe1dab1b30f7c5c7e581b93-Abstract.html}
}

@inproceedings{gshard,
  author    = {Dmitry Lepikhin and HyoukJoong Lee and Yuanzhong Xu and
               Dehao Chen and Orhan Firat and Yanping Huang and Maxim Krikun and
               Noam Shazeer and Zhifeng Chen},
  title     = {{GShard}: Scaling Giant Models with Conditional Computation
               and Automatic Sharding},
  booktitle = {The Ninth International Conference on Learning Representations
               (ICLR)},
  year      = {2021},
  url       = {https://openreview.net/forum?id=qrwe7XHTmYb}
}

@inproceedings{alpa,
  author    = {Lianmin Zheng and Zhuohan Li and Hao Zhang and Yonghao Zhuang and
               Zhifeng Chen and Yanping Huang and Yida Wang and Yuanzhong Xu and
               Danyang Zhuo and Eric P. Xing and Joseph E. Gonzalez and Ion Stoica},
  title     = {Alpa: Automating Inter- and Intra-Operator Parallelism for
               Distributed Deep Learning},
  booktitle = {Proceedings of the 16th USENIX Symposium on Operating Systems
               Design and Implementation (OSDI)},
  publisher = {USENIX Association},
  address   = {Carlsbad, CA},
  pages     = {559--578},
  year      = {2022},
  url       = {https://www.usenix.org/conference/osdi22/presentation/zheng-lianmin}
}

@misc{gspmd,
  author    = {Yuanzhong Xu and HyoukJoong Lee and Dehao Chen and Blake Hechtman and
               Yanping Huang and Rahul Joshi and Maxim Krikun and Dmitry Lepikhin and
               Andy Ly and Marcello Maggioni and Ruoming Pang and Noam Shazeer and
               Shibo Wang and Tao Wang and Yonghui Wu and Zhifeng Chen},
  title     = {{GSPMD}: General and Scalable Parallelization for {ML}
               Computation Graphs},
  archivePrefix = {arXiv},
  eprint    = {2105.04663},
  year      = {2021},
  doi       = {10.48550/arXiv.2105.04663},
  url       = {https://arxiv.org/abs/2105.04663}
}

@inproceedings{sequence-parallel,
  author    = {Vijay Anand Korthikanti and Jared Casper and Sangkug Lym and
               Lawrence McAfee and Michael Andersch and Mohammad Shoeybi and
               Bryan Catanzaro},
  title     = {Reducing Activation Recomputation in Large Transformer Models},
  booktitle = {Proceedings of Machine Learning and Systems},
  volume    = {5},
  publisher = {MLSys},
  pages     = {341--353},
  year      = {2023},
  url       = {https://proceedings.mlsys.org/paper_files/paper/2023/hash/80083951326cf5b35e5100260d64ed81-Abstract-mlsys2023.html}
}

@inproceedings{pytorch,
  author    = {Adam Paszke and Sam Gross and Francisco Massa and Adam Lerer and
               James Bradbury and Gregory Chanan and Trevor Killeen and
               Zeming Lin and Natalia Gimelshein and Luca Antiga and
               Alban Desmaison and Andreas K{\"o}pf and Edward Yang and
               Zachary DeVito and Martin Raison and Alykhan Tejani and
               Sasank Chilamkurthy and Benoit Steiner and Lu Fang and
               Junjie Bai and Soumith Chintala},
  title     = {{PyTorch}: An Imperative Style, High-Performance Deep Learning
               Library},
  booktitle = {Advances in Neural Information Processing Systems},
  volume    = {32},
  publisher = {Curran Associates, Inc.},
  year      = {2019},
  url       = {https://proceedings.neurips.cc/paper/2019/hash/bdbca288fee7f92f2bfa9f7012727740-Abstract.html}
}

@inproceedings{volkov-gemm,
  author    = {Vasily Volkov and James W. Demmel},
  title     = {Benchmarking {GPUs} to Tune Dense Linear Algebra},
  booktitle = {Proceedings of the 2008 ACM/IEEE Conference on Supercomputing
               (SC)},
  publisher = {IEEE},
  pages     = {1--11},
  year      = {2008},
  doi       = {10.1109/SC.2008.5214359},
  url       = {https://doi.org/10.1109/SC.2008.5214359}
}

@inproceedings{coconet,
  author    = {Abhinav Jangda and Jun Huang and Guodong Liu and
               Amir Hossein Nodehi Sabet and Saeed Maleki and Youshan Miao and
               Madanlal Musuvathi and Todd Mytkowicz and Olli Saarikivi},
  title     = {Breaking the Computation and Communication Abstraction Barrier
               in Distributed Machine Learning Workloads},
  booktitle = {Proceedings of the 27th ACM International Conference on
               Architectural Support for Programming Languages and Operating
               Systems (ASPLOS)},
  pages     = {402--416},
  publisher = {Association for Computing Machinery},
  address   = {New York, NY, USA},
  year      = {2022},
  doi       = {10.1145/3503222.3507778},
  url       = {https://doi.org/10.1145/3503222.3507778}
}

@inproceedings{decomposition-asplos23,
  author    = {Shibo Wang and Jinliang Wei and Amit Sabne and Andy Davis and
               Berkin Ilbeyi and Blake Hechtman and Dehao Chen and
               Karthik Srinivasa Murthy and Marcello Maggioni and Qiao Zhang and
               Sameer Kumar and Tongfei Guo and Yuanzhong Xu and Zongwei Zhou},
  title     = {Overlap Communication with Dependent Computation via
               Decomposition in Large Deep Learning Models},
  booktitle = {Proceedings of the 28th ACM International Conference on
               Architectural Support for Programming Languages and Operating
               Systems (ASPLOS), Volume 1},
  pages     = {93--106},
  publisher = {Association for Computing Machinery},
  address   = {New York, NY, USA},
  year      = {2022},
  doi       = {10.1145/3567955.3567959},
  url       = {https://doi.org/10.1145/3567955.3567959}
}

@inproceedings{megatron-sc21,
  author    = {Deepak Narayanan and Mohammad Shoeybi and Jared Casper and
               Patrick LeGresley and Mostofa Patwary and Vijay Korthikanti and
               Dmitri Vainbrand and Prethvi Kashinkunti and Julie Bernauer and
               Bryan Catanzaro and Amar Phanishayee and Matei Zaharia},
  title     = {Efficient Large-Scale Language Model Training on {GPU} Clusters
               Using {Megatron-LM}},
  booktitle = {Proceedings of the International Conference for High Performance
               Computing, Networking, Storage and Analysis (SC)},
  pages     = {1--15},
  publisher = {Association for Computing Machinery},
  address   = {New York, NY, USA},
  year      = {2021},
  doi       = {10.1145/3458817.3476209},
  url       = {https://doi.org/10.1145/3458817.3476209}
}

@inproceedings{centauri,
  author    = {Chang Chen and Xiuhong Li and Qianchao Zhu and Jiangfei Duan and
               Peng Sun and Xingcheng Zhang and Chao Yang},
  title     = {Centauri: Enabling Efficient Scheduling for
               Communication-Computation Overlap in Large Model Training via
               Communication Partitioning},
  booktitle = {Proceedings of the 29th ACM International Conference on
               Architectural Support for Programming Languages and Operating
               Systems (ASPLOS), Volume 3},
  pages     = {178--191},
  publisher = {Association for Computing Machinery},
  address   = {New York, NY, USA},
  year      = {2024},
  doi       = {10.1145/3620666.3651379},
  url       = {https://doi.org/10.1145/3620666.3651379}
}

@inproceedings{syndicate,
  author    = {Kshiteej Mahajan and Ching-Hsiang Chu and
               Srinivas Sridharan and Aditya Akella},
  title     = {Better Together: Jointly Optimizing {ML} Collective Scheduling
               and Execution Planning Using {SYNDICATE}},
  booktitle = {Proceedings of the 20th USENIX Symposium on Networked Systems
               Design and Implementation (NSDI)},
  pages     = {809--824},
  publisher = {USENIX Association},
  address   = {Boston, MA},
  year      = {2023},
  url       = {https://www.usenix.org/conference/nsdi23/presentation/mahajan}
}

@techreport{domino,
  author      = {Guanhua Wang and Chengming Zhang and Zheyu Shen and Ang Li and
                 Olatunji Ruwase},
  title       = {Domino: Eliminating Communication in {LLM} Training via Generic
                 Tensor Slicing and Overlapping},
  institution = {Microsoft Research},
  type        = {Technical Report},
  number      = {MSR-TR-2024-40},
  month       = sep,
  year        = {2024},
  url         = {https://www.microsoft.com/en-us/research/publication/domino-eliminating-communication-in-llm-training-via-generic-tensor-slicing-and-overlapping/}
}

@inproceedings{torchtitan,
  author    = {Wanchao Liang and Tianyu Liu and Less Wright and Will Constable and
               Andrew Gu and Chien-Chin Huang and Iris Zhang and Wei Feng and
               Howard Huang and Junjie Wang and Sanket Purandare and
               Gokul Nadathur and Stratos Idreos},
  title     = {TorchTitan: One-Stop PyTorch Native Solution for Production Ready
               {LLM} Pretraining},
  booktitle = {The Thirteenth International Conference on Learning
               Representations (ICLR)},
  year      = {2025},
  url       = {https://openreview.net/forum?id=SFN6Wm7YBI}
}

@misc{asynctp,
  author       = {Horace He and Less Wright and Luca Wehrstedt and Tianyu Liu and
                  Wanchao Liang},
  title        = {[Distributed w/ {TorchTitan}] Introducing Async Tensor Parallelism
                  in {PyTorch}},
  year         = {2024},
  howpublished = {PyTorch Developer Forums},
  url          = {https://discuss.pytorch.org/t/distributed-w-torchtitan-introducing-async-tensor-parallelism-in-pytorch/209487},
  note         = {Accessed 6 September 2026}
}

@misc{flux,
  author  = {Li-Wen Chang and Wenlei Bao and Qi Hou and Chengquan Jiang and
             Ningxin Zheng and Yinmin Zhong and Xuanrun Zhang and Zuquan Song and
             Chengji Yao and Ziheng Jiang and Haibin Lin and Xin Jin and Xin Liu},
  title   = {{FLUX}: Fast Software-Based Communication Overlap On {GPUs} Through
             Kernel Fusion},
  archivePrefix = {arXiv},
  eprint  = {2406.06858},
  year    = {2024},
  doi     = {10.48550/arXiv.2406.06858},
  url     = {https://arxiv.org/abs/2406.06858}
}

@inproceedings{fused-collective,
  author    = {Kishore Punniyamurthy and Khaled Hamidouche and
               Bradford M. Beckmann},
  title     = {Optimizing Distributed {ML} Communication with Fused
               Computation-Collective Operations},
  booktitle = {Proceedings of the International Conference for High Performance
               Computing, Networking, Storage and Analysis (SC)},
  pages     = {1--17},
  publisher = {IEEE},
  year      = {2024},
  doi       = {10.1109/SC41406.2024.00094},
  url       = {https://doi.org/10.1109/SC41406.2024.00094}
}

@inproceedings{tilelink,
  author    = {Size Zheng and Jin Fang and Xuegui Zheng and Qi Hou and
               Wenlei Bao and Ningxin Zheng and Ziheng Jiang and Dongyang Wang and
               Jianxi Ye and Haibin Lin and Li-Wen Chang and Xin Liu},
  title     = {TileLink: Generating Efficient Compute-Communication Overlapping
               Kernels Using Tile-Centric Primitives},
  booktitle = {Proceedings of Machine Learning and Systems},
  volume    = {7},
  publisher = {MLSys},
  year      = {2025},
  url       = {https://proceedings.mlsys.org/paper_files/paper/2025/hash/c6ee784cbe46d854843e4c883a3321ef-Abstract-Conference.html}
}

@inproceedings{syncopate,
  author    = {Xinwei Qiang and Yue Guan and Zhengding Hu and Keren Zhou and
               Yufei Ding and Adnan Aziz},
  title     = {Syncopate: Efficient Multi-{GPU} {AI} Kernels via Automatic
               Chunk-Centric Compute-Communication Overlap},
  booktitle = {Proceedings of the 20th USENIX Symposium on Operating Systems
               Design and Implementation (OSDI)},
  pages     = {331--347},
  publisher = {USENIX Association},
  address   = {Seattle, WA},
  year      = {2026},
  url       = {https://www.usenix.org/conference/osdi26/presentation/qiang}
}

@inproceedings{flashoverlap,
  author    = {Ke Hong and Xiuhong Li and Minxu Liu and Qiuli Mao and
               Tianqi Wu and Zixiao Huang and Lufang Chen and Zhong Wang and
               Yichong Zhang and Zhenhua Zhu and Guohao Dai and Yu Wang},
  title     = {Efficient and Adaptable Overlapping for Computation and
               Communication via Signaling and Reordering},
  booktitle = {Proceedings of the 21st European Conference on Computer Systems
               (EuroSys)},
  pages     = {1894--1911},
  publisher = {Association for Computing Machinery},
  address   = {New York, NY, USA},
  year      = {2026},
  doi       = {10.1145/3767295.3769370},
  url       = {https://doi.org/10.1145/3767295.3769370}
}

@inproceedings{t3,
  author    = {Suchita Pati and Shaizeen Aga and Mahzabeen Islam and
               Nuwan Jayasena and Matthew D. Sinclair},
  title     = {{T3}: Transparent Tracking \& Triggering for Fine-grained Overlap
               of Compute \& Collectives},
  booktitle = {Proceedings of the 29th ACM International Conference on
               Architectural Support for Programming Languages and Operating
               Systems (ASPLOS), Volume 2},
  pages     = {1146--1164},
  publisher = {Association for Computing Machinery},
  address   = {New York, NY, USA},
  year      = {2024},
  doi       = {10.1145/3620665.3640410},
  url       = {https://doi.org/10.1145/3620665.3640410}
}

@misc{moe-tile-signaling,
  author  = {Minyu Cui and Anna Wingkvist and Morgan Ericsson},
  title   = {Fine-Grained Computation-Communication Overlap via Tile-Level
             Signaling and Scheduling for Mixture-of-Experts},
  archivePrefix = {arXiv},
  eprint  = {2607.19539},
  note    = {Accepted at the 55th International Conference on Parallel
             Processing (ICPP 2026); arXiv version cited},
  year    = {2026},
  doi     = {10.48550/arXiv.2607.19539},
  url     = {https://arxiv.org/abs/2607.19539}
}

@inproceedings{gpu-maestro,
  author    = {Jason Jong Kyu Park and Yongjun Park and Scott Mahlke},
  title     = {Dynamic Resource Management for Efficient Utilization of
               Multitasking {GPUs}},
  booktitle = {Proceedings of the 22nd International Conference on Architectural
               Support for Programming Languages and Operating Systems (ASPLOS)},
  publisher = {Association for Computing Machinery},
  address   = {New York, NY, USA},
  pages     = {527--540},
  year      = {2017},
  doi       = {10.1145/3037697.3037707},
  url       = {https://doi.org/10.1145/3037697.3037707}
}

@inproceedings{hfuse,
  author    = {Ao Li and Bojian Zheng and Gennady Pekhimenko and Fan Long},
  title     = {Automatic Horizontal Fusion for {GPU} Kernels},
  booktitle = {Proceedings of the 2022 IEEE/ACM International Symposium on
               Code Generation and Optimization (CGO)},
  pages     = {14--27},
  publisher = {IEEE},
  year      = {2022},
  doi       = {10.1109/CGO53902.2022.9741270},
  url       = {https://doi.org/10.1109/CGO53902.2022.9741270}
}

@inproceedings{comet,
  author    = {Shulai Zhang and Ningxin Zheng and Haibin Lin and Ziheng Jiang and
               Wenlei Bao and Chengquan Jiang and Qi Hou and Weihao Cui and
               Size Zheng and Li-Wen Chang and Quan Chen and Xin Liu},
  title     = {{COMET}: Fine-Grained Computation-Communication Overlapping for
               Mixture-of-Experts},
  booktitle = {Proceedings of Machine Learning and Systems},
  volume    = {7},
  publisher = {MLSys},
  year      = {2025},
  url       = {https://proceedings.mlsys.org/paper_files/paper/2025/hash/e27ea0cd50b798ff8942caf9203f0992-Abstract-Conference.html}
}

@inproceedings{parallelkittens,
  author    = {Stuart H. Sul and Simran Arora and Benjamin F. Spector and
               Christopher R{\'e}},
  title     = {ParallelKittens: Systematic and Practical Simplification of
               Multi-{GPU} {AI} Kernels},
  booktitle = {Proceedings of Machine Learning and Systems},
  volume    = {8},
  pages     = {1076--1089},
  publisher = {MLSys},
  year      = {2026},
  url       = {https://proceedings.mlsys.org/paper_files/paper/2026/hash/ff997469ac66cf893c4183efeb22212a-Abstract-Conference.html}
}

@misc{resource-aware-overlap,
  author  = {Minyu Cui and Miquel Peric{\`a}s},
  title   = {Resource-Aware Computation-Communication Overlap for Multi-{GPU}
             {ML} Workloads},
  archivePrefix = {arXiv},
  eprint  = {2606.09200},
  note    = {Accepted at the AI on HPC Workshop at ISC 2026;
             arXiv version cited},
  year    = {2026},
  doi     = {10.48550/arXiv.2606.09200},
  url     = {https://arxiv.org/abs/2606.09200}
}

@misc{ficco,
  author  = {Shagnik Pal and Shaizeen Aga and Suchita Pati and
             Mahzabeen Islam and Lizy K. John},
  title   = {Design Space Exploration of {DMA} Based Finer-Grain Compute
             Communication Overlap},
  archivePrefix = {arXiv},
  eprint  = {2512.10236},
  year    = {2025},
  doi     = {10.48550/arXiv.2512.10236},
  url     = {https://arxiv.org/abs/2512.10236}
}

@inproceedings{ace,
  author    = {Saeed Rashidi and Matthew Denton and Srinivas Sridharan and
               Sudarshan Srinivasan and Amoghavarsha Suresh and Jade Nie and
               Tushar Krishna},
  title     = {Enabling Compute-Communication Overlap in Distributed Deep
               Learning Training Platforms},
  booktitle = {Proceedings of the 48th Annual International Symposium on
               Computer Architecture (ISCA)},
  pages     = {540--553},
  publisher = {IEEE},
  year      = {2021},
  doi       = {10.1109/ISCA52012.2021.00049},
  url       = {https://doi.org/10.1109/ISCA52012.2021.00049}
}

@inproceedings{ark,
  author    = {Changho Hwang and KyoungSoo Park and Ran Shu and Xinyuan Qu and
               Peng Cheng and Yongqiang Xiong},
  title     = {{ARK}: {GPU}-Driven Code Execution for Distributed Deep
               Learning},
  booktitle = {Proceedings of the 20th USENIX Symposium on Networked Systems
               Design and Implementation (NSDI)},
  pages     = {87--101},
  publisher = {USENIX Association},
  address   = {Boston, MA},
  year      = {2023},
  url       = {https://www.usenix.org/conference/nsdi23/presentation/hwang}
}

@inproceedings{ptx-memory-model,
  author    = {Daniel Lustig and Sameer Sahasrabuddhe and Olivier Giroux},
  title     = {A Formal Analysis of the {NVIDIA PTX} Memory Consistency Model},
  booktitle = {Proceedings of the 24th International Conference on Architectural
               Support for Programming Languages and Operating Systems (ASPLOS)},
  publisher = {Association for Computing Machinery},
  address   = {New York, NY, USA},
  pages     = {257--270},
  year      = {2019},
  doi       = {10.1145/3297858.3304043},
  url       = {https://doi.org/10.1145/3297858.3304043}
}

@inproceedings{evt,
  author    = {Zhaodong Chen and Andrew Kerr and Richard Cai and
               Jack Kosaian and Haicheng Wu and Yufei Ding and Yuan Xie},
  title     = {{EVT}: Accelerating Deep Learning Training with Epilogue
               Visitor Tree},
  booktitle = {Proceedings of the 29th ACM International Conference on
               Architectural Support for Programming Languages and Operating
               Systems (ASPLOS), Volume 3},
  pages     = {301--316},
  publisher = {Association for Computing Machinery},
  address   = {New York, NY, USA},
  year      = {2024},
  doi       = {10.1145/3620666.3651369},
  url       = {https://doi.org/10.1145/3620666.3651369}
}

@misc{cutlass,
  author       = {{NVIDIA}},
  title        = {{CUTLASS}: {CUDA} Templates for Linear Algebra Subroutines},
  year         = {2026},
  howpublished = {Software, commit 08185b9c},
  url          = {https://github.com/NVIDIA/cutlass/tree/08185b9c}
}

@misc{cuda-programming-guide,
  author       = {{NVIDIA}},
  title        = {{CUDA} Programming Guide},
  year         = {2026},
  howpublished = {NVIDIA Documentation},
  url          = {https://docs.nvidia.com/cuda/cuda-programming-guide/index.html},
  note         = {Accessed 3 September 2026}
}

@inproceedings{sccl,
  author    = {Zixian Cai and Zhengyang Liu and Saeed Maleki and Madanlal Musuvathi and
               Todd Mytkowicz and Jacob Nelson and Olli Saarikivi},
  title     = {Synthesizing Optimal Collective Algorithms},
  booktitle = {Proceedings of the 26th ACM SIGPLAN Symposium on Principles and
               Practice of Parallel Programming (PPoPP)},
  publisher = {Association for Computing Machinery},
  address   = {New York, NY, USA},
  pages     = {62--75},
  year      = {2021},
  doi       = {10.1145/3437801.3441620},
  url       = {https://doi.org/10.1145/3437801.3441620}
}

@inproceedings{taccl,
  author    = {Aashaka Shah and Vijay Chidambaram and Meghan Cowan and Saeed Maleki and
               Madan Musuvathi and Todd Mytkowicz and Jacob Nelson and Olli Saarikivi and
               Rachee Singh},
  title     = {{TACCL}: Guiding Collective Algorithm Synthesis Using
               Communication Sketches},
  booktitle = {Proceedings of the 20th USENIX Symposium on Networked Systems
               Design and Implementation (NSDI)},
  publisher = {USENIX Association},
  address   = {Boston, MA},
  pages     = {593--612},
  year      = {2023},
  url       = {https://www.usenix.org/conference/nsdi23/presentation/shah}
}

@inproceedings{mscclang,
  author    = {Meghan Cowan and Saeed Maleki and Madanlal Musuvathi and
               Olli Saarikivi and Yifan Xiong},
  title     = {{MSCCLang}: {Microsoft} Collective Communication Language},
  booktitle = {Proceedings of the 28th ACM International Conference on
               Architectural Support for Programming Languages and Operating
               Systems (ASPLOS), Volume 2},
  pages     = {502--514},
  publisher = {Association for Computing Machinery},
  address   = {New York, NY, USA},
  year      = {2023},
  doi       = {10.1145/3575693.3575724},
  url       = {https://doi.org/10.1145/3575693.3575724}
}

@misc{cublas,
  author       = {{NVIDIA}},
  title        = {{cuBLAS} Library},
  year         = {2026},
  howpublished = {NVIDIA Documentation},
  url          = {https://docs.nvidia.com/cuda/cublas/},
  note         = {Accessed 3 September 2026}
}

@misc{nccl,
  author       = {{NVIDIA}},
  title        = {{NVIDIA} Collective Communications Library ({NCCL})},
  year         = {2026},
  howpublished = {NVIDIA Developer},
  url          = {https://developer.nvidia.com/nccl},
  note         = {Undated web page; 2026 denotes the version accessed on
                  9 September 2026}
}

@misc{nvshmem,
  author       = {{NVIDIA}},
  title        = {{NVSHMEM}},
  year         = {2026},
  howpublished = {NVIDIA Developer},
  url          = {https://developer.nvidia.com/nvshmem},
  note         = {Undated web page; 2026 denotes the version accessed on
                  9 September 2026}
}

@inproceedings{mscclpp,
  author    = {Changho Hwang and Peng Cheng and Roshan Dathathri and
               Abhinav Jangda and Saeed Maleki and Madan Musuvathi and
               Olli Saarikivi and Aashaka Shah and Ziyue Yang and Binyang Li and
               Caio Rocha and Qinghua Zhou and Mahdieh Ghazimirsaeed and
               Sreevatsa Anantharamu and Jithin Jose},
  title     = {{MSCCL++}: Rethinking {GPU} Communication Abstractions for
               {AI} Inference},
  booktitle = {Proceedings of the 31st ACM International Conference on
               Architectural Support for Programming Languages and Operating
               Systems (ASPLOS), Volume 2},
  pages     = {1201--1215},
  publisher = {Association for Computing Machinery},
  address   = {New York, NY, USA},
  year      = {2026},
  doi       = {10.1145/3779212.3790188},
  url       = {https://doi.org/10.1145/3779212.3790188}
}

@misc{triton-distributed,
  author  = {Size Zheng and Wenlei Bao and Qi Hou and Xuegui Zheng and
             Jin Fang and Chenhui Huang and Tianqi Li and Haojie Duanmu and
             Renze Chen and Ruifan Xu and Yifan Guo and Ningxin Zheng and
             Ziheng Jiang and Xinyi Di and Dongyang Wang and Jianxi Ye and
             Haibin Lin and Li-Wen Chang and Liqiang Lu and Yun Liang and
             Jidong Zhai and Xin Liu},
  title   = {Triton-Distributed: Programming Overlapping Kernels on Distributed
             {AI} Systems with the {Triton} Compiler},
  archivePrefix = {arXiv},
  eprint  = {2504.19442},
  year    = {2025},
  doi     = {10.48550/arXiv.2504.19442},
  url     = {https://arxiv.org/abs/2504.19442}
}

@inproceedings{triton,
  author    = {Philippe Tillet and H. T. Kung and David Cox},
  title     = {Triton: An Intermediate Language and Compiler for Tiled Neural
               Network Computations},
  booktitle = {Proceedings of the 3rd ACM SIGPLAN International Workshop on
               Machine Learning and Programming Languages (MAPL)},
  pages     = {10--19},
  publisher = {Association for Computing Machinery},
  address   = {New York, NY, USA},
  year      = {2019},
  doi       = {10.1145/3315508.3329973},
  url       = {https://doi.org/10.1145/3315508.3329973}
}

\appendix

\section{Communication Bandwidth in Isolation}
\label{sec:appendix:isolated}

\begin{figure}[H]
  \centering
  \includegraphics[width=\columnwidth]{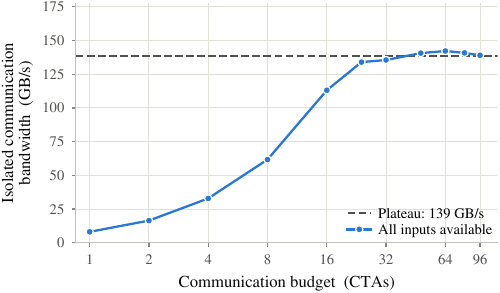}
  \caption{Communication bandwidth with all inputs available and no competing
  producer. Figure~\ref{fig:resource-coupling} shows execution with
  concurrent computation.}
  \Description{Bandwidth rises steeply with the communication budget
  over the first several settings, then varies within a narrow band around a
  fitted plateau.}
  \label{fig:isolated-response}
\end{figure}

\section{Full Communication-Budget Sweep}
\label{sec:appendix:budget}

\begin{figure*}[t]
  \centering
  \includegraphics[width=\textwidth]{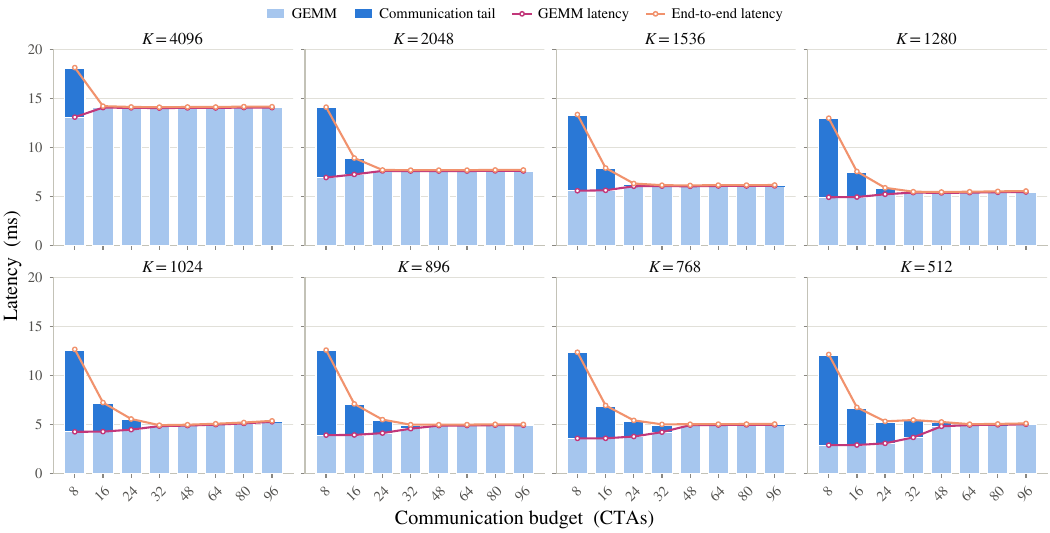}
  \caption{Coupled execution across communication budgets.
  Budgets cap concurrent consumer work. Bars show computation windows and
  communication tails; curves show GEMM and end-to-end latency, including
  completion overhead. Panels vary local $K$ with $M$ and $N$ fixed.
  Figure~\ref{fig:resource-coupling} highlights the $K=768$ case.}
  \Description{Across eight values of K, increasing the budget initially
  shortens the communication tail. Deeper reductions reach a broad latency
  plateau; for shallower reductions, producer delay can offset the benefit
  of further tail reduction.}
  \label{fig:resource-coupling-full}
\end{figure*}

Figure~\ref{fig:resource-coupling-full} gives the full sweep discussed in
\S\ref{sec:eval:coupling}, including the two cases in
Figure~\ref{fig:motivation} and the intermediate case in
Figure~\ref{fig:resource-coupling}.

\section{Overlap Efficiency Across Workloads}
\label{sec:appendix:efficiency}

\begin{figure*}[t]
  \centering
  \includegraphics[width=\textwidth]{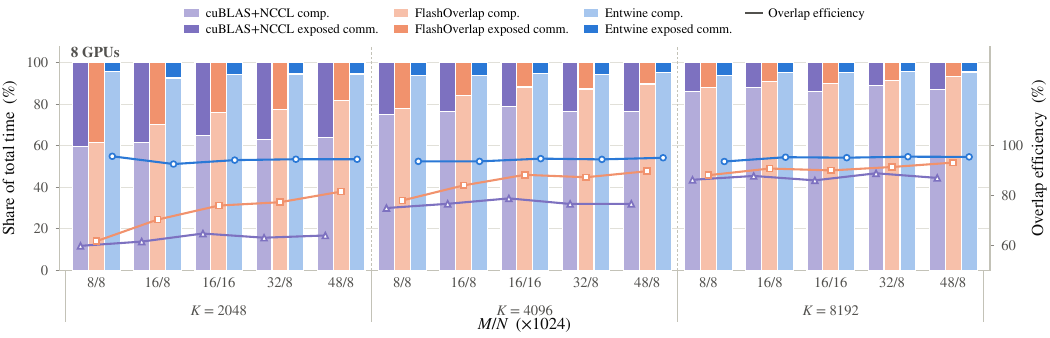}
  \caption{Overlap efficiency.
  On the left axis, light bar segments show overlap efficiency.
  Dark segments show the remaining fraction.
  Curves repeat overlap efficiency on the right axis to show its trend;
  the two axes use different ranges. Each method uses its own uncontended
  producer as the computation baseline.}
  \Description{Grouped stacked bars and overlaid curves show overlap
  efficiency for three methods across 15 workloads, grouped by local K.
  The curves encode the same fractions as the light bar segments, using
  the right axis. Dark segments are the complementary fractions.}
  \label{fig:overlap-efficiency}
\end{figure*}

Figure~\ref{fig:overlap-efficiency} supplements the aggregate overlap
efficiency in \S\ref{sec:eval:operator} with per-workload results.
For each method, overlap efficiency is the ratio of uncontended producer
time to end-to-end operator time. The remaining fraction includes the cost
of communication and any producer slowdown relative to that baseline.
It is therefore distinct from the
measured communication tail in Figure~\ref{fig:resource-coupling-full}.

On all 15 workloads, \systemname{} achieves higher overlap efficiency than
cuBLAS~+~NCCL and FlashOverlap. Its efficiency remains high across the
three reduction dimensions, while both baselines improve more substantially
as $K$ increases.

\vfuzz=2pt

\end{document}